\documentclass[aps,prl,twocolumn,showpacs]{revtex4}
\usepackage{graphicx}

\newcommand{\eqb}{\begin{equation}}
\newcommand{\eqe}{\end{equation}}
\newcommand{\arrb}{\begin{eqnarray}}
\newcommand{\arre}{\end{eqnarray}}

\def\threej #1.#2.#3.#4.#5.#6. {\pmatrix{#1 & #2 & #3 \cr #4 & #5 & #6 \cr}}

\begin{document}

\title{Experimental verification of minima in excited long-range Rydberg states of
Rb$_{2}$}
\author{Chris H. Greene$^{1}$}
\author{Edward L. Hamilton$^{2}$}
\author{Heather Crowell$^{3}$} 
\author{Cedomil Vadla$^{4}$}
\author{Kay Niemax$^{5}$}

\affiliation{$^{1}$Department of Physics and JILA, University of Colorado, Boulder,
CO 80309-0440} 
\affiliation{$^{2}$Department of Chemistry, Northwestern University, Evanston,
IL 60208-3113} 
\affiliation{$^{3}$Department of Chemistry and JILA, University of Colorado, Boulder,
CO 80309-0440} 

\affiliation{$^{4}$Institute of Physics, Zagreb, Croatia} 
\affiliation{$^{5}$ISAS -- Institute for Analytical Sciences, Dortmund, Germany} 

\date{\today}

\begin{abstract}
Recent theoretical studies with alkali atoms A$^{\ast }$
excited to high Rydberg states predicted the existence of ultra long-range molecular
bound states.
Such excited dimers have large electric dipole moments
which, in combination with their long radiative lifetimes, make them
excellent candidates for manipulation in applications.
This letter reports on experimental investigations of the self-broadening of Rb principal series lines, which revealed multiple satellites in the line wings. The positions of the satellites agree quantitatively with theoretically-predicted minima in the excited long-range Rydberg states of Rb$_2$. 
\end{abstract}







\maketitle

%

Two different types of bound Rydberg molecules have been 
proposed in recent years as intriguing candidates
for possible study. In the first case 
\cite{Greene2000,Granger2001,Chibisov2002,Hamilton2002,Khuskivadze2002}, 
a bound state is
created by the interaction between an excited atom (A$^{\ast }$) and a
ground state atom (A). The properties of low-energy scattering of the
quasi-free electron of the Rydberg atom A$^{\ast }$ on the 
perturber A create the attraction needed for dimer formation. 
These Rydberg molecules are comparable to
the size of the excited atoms, and the corresponding electronic wave
functions are characterized by specific spatial forms, two of which are the
``trilobite'' and ``butterfly'' states. These molecules are also predicted
to be polar diatomic molecules possessing dipole moments (even the
homonuclear species) that are huge compared to typical polar diatomic
molecules. This property makes them easy to manipulate by even very small
fields. A growing body of theoretical evidence strongly suggests the
existence of these unusual molecular states, whose Born-Oppenheimer
potential curves oscillate like a wavefunction (see, e.g., Refs.
\cite{Valiron1984,dePrunele1987,DuGreene1987,DuGreene1987E, LiuRost2006})
But until now, their existence has not been confirmed experimentally.

The second type of bound molecular Rydberg state is predicted to be created
by the interaction of two Rydberg atoms A$^{\ast }+$ A$^{\ast }$ at very 
large distances $R$\cite{Boisseau2002}. Here, the potential
energy curve can be exclusively described by the long-range electrostatic
interaction as a sum of $C_n/R^n$ terms.  
The binding energies of both types of proposed long-range
Rydberg molecules are very small in comparison with thermal kinetic energies
at room temperatures. However, their formation and detection was predicted to
be possible in ultra-cold gases or Bose-Einstein condensates. 
Laser spectroscopic experiments performed in ultra-cold gases (\cite
{deOliveira2003,Farooqi2003,Singer2004} and references therein)
have provided evidence for the existence of bound molecular Rydberg states
of the A$^{\ast }+$A$^{\ast }$ type \cite{deOliveira2003,Singer2004}.

Our letter is focused on the experimental verification of the 
oscillatory bound state potentials 
in the Rb$^{\ast }+$Rb system associated with electron-atom scattering
resonances. In contrast to experiments mentioned above, we apply the 
well-established method of spectral line wing absorption measurement under
thermal conditions\cite{Szudy1996}.  Such measurements reveal satellites
in the wings of collisionally-broadened lines at extrema in difference
potentials of the relevant molecular states.  The positions, shapes,
and intensities of satellites depend on the difference potentials as 
functions of the interatomic separation $R$. 
Therefore, absorption measurements on thermal Rb
vapor have been performed to study the line wings of the collisionally
broadened principal series lines 5S--nP. We were encouraged to make such
measurements since peculiar satellite structures have been observed in the
line wings of collisionally broadened principal series lines of Cs, 
dating back to more than
30 years ago \cite{Eliseev1971,Niemax1972}. The absolute satellite strengths
in terms of reduced absorption coefficients were published later for the
6S--nP (n=8--13) lines \cite{Heinke1983}. Similar satellite features have
also been observed in Rb spectra (see remark in \cite{Heinke1983}). However,
these results were never published, primarily due to our poor
understanding of the physical origin of these regular satellite features, 
through conventional line-broadening theory.

The long history of investigations of far wing alkali line broadening
resulted, for the first resonance line of each species, in complete
agreement between theory and experiment (see, e.g., \cite{Gallagher2006}). 
Regarding the theory, there is a smooth transition between \textit{ab
initio} calculations and the perturbation approach using the long range
multipole interactions. Recent \textit{ab initio} calculations also yielded
a satisfactory explanation of the measured structure of the second principal
series line of Rb, the 5S--6P transition \cite{Ban2004}. However, for the
transitions to the third and higher resonance states neither \textit{ab
initio} nor the usual perturbation calculations can provide any explanation
for the pronounced satellite structures in the quasistatic region of the
line wings. Different models proposed and
discussions held at conferences dealing with spectral line shapes have not
to date produced satisfactory explanations. Moreover, the unexpected
oscillations of experimental line shift and impact broadening parameters, in
their dependence on the effective quantum number for alkali Rydberg line
series, are still without adequate explanation (see \cite{Heinke1983} and
references therein).

The experiments were performed with a Rb vapor-filled heat pipe, using
the classical white light absorption method with a tungsten-halogen lamp and
a 0.75-m Czerny-Turner type monochromator supplied with a EMI 9789QA
photomultiplier. Although the experimental arrangement is in principle very
simple, the realization of the experiment required solving some technical
problems connected with the production of a stable rubidium vapor at
pressures of about 100 mbar (number density $\mathrm{N_{Rb}}\geq 10^{18}\,%
\text{cm}^{-3}$). This was necessary to get stable and measurable absorption
in the line wings. The specific details of our experimental arrangement are
given in \cite{VadlaBeuc2006} and will be repeated here only briefly. The rubidium vapor
column was generated in the middle part of a
stainless-steel heat pipe by the use of an outer
electric heater. Argon
was used as a buffer gas to protect the quartz windows at the cold ends of
the heat pipe oven from the corrosive influence
of the hot rubidium vapor. The
heat pipe was running in the heat pipe mode, i.e., the rubidium pressure was
equal to the buffer gas pressure. It was possible to achieve a rubidium
vapor pressure of about 200 mbar. However, it is well known that strong 
turbulent ''fogs''
(particle clouds) appear in the transition regions between the hot vapor and
the cold buffer gas at vapor pressures larger than about 30 mbar. 
As shown in \cite
{VadlaBeuc2006}, this effect can be efficiently removed with a help of an
additional rod heater built in along the heat pipe axis which accomplished
stable rubidium vapor conditions by overheating the vapor. E.g., for the
maximum temperature of $\sim \!1100$ K near to the central heater, the metal
bath temperature at the heat pipe wall was $\sim \!800$ K.

The absolute reduced absorption coefficients $k_{R}=k/N^{2}$ of the
measured lines were obtained using the narrow triplet molecular band
at 605 nm as a reference for $N$ determination.  The absorption of this 
molecular band was calibrated using the far-wing absorption of the first
resonance lines\cite{Horvatic1993}, in the same way as for the 
corresponding potassium triplet band\cite{Heinke1983}.  To avoid spectral
interference of the line wings with the strongly temperature-dependent 
Rb dimer spectrum, we made spatially-resolved absorption 
measurements in the overheated vapor.
The spatially resolved absorption
spectra were related to measurements in thin absorption columns 
at various distances from the heat pipe axis, i.e., in
different temperature zones. By this the dissociation of the ground state
dimers could be varied and dimer interferences with the atomic line shapes
could be excluded. The reduced absorption coefficients of the principal
series lines given below were found to be constant in the applied
temperature range.
\begin{figure}[h]
\begin{center}
\includegraphics[width=3.7in]{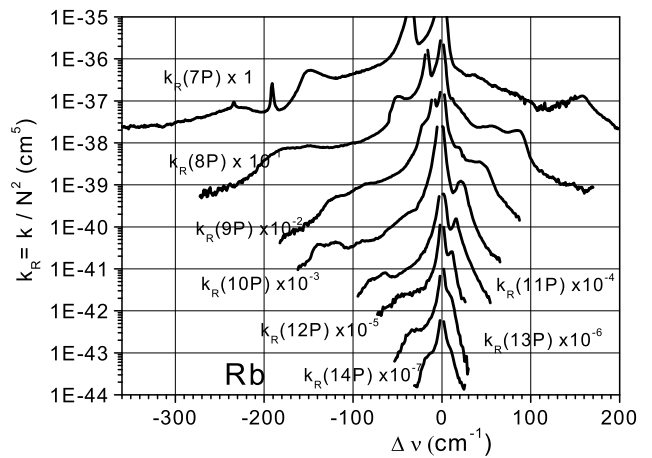}
\end{center}
\caption{The reduced absorption coefficients $k_{R}(\text{nP})=k(\text{nP}%
)/N_{\mathrm{Rb}}^{2}$ of the quasistatic wings of some Rb 5S--nP lines are
plotted versus the wavenumber difference from the respective line
centers. For better representation, the individual spectra have been
multiplied by different factors of 10 as indicated.}%
\label{abscoeff}
\end{figure}

The measured reduced absorption coefficients $k_{R} = k/N^{2}$ of some
principal series line are shown in Fig. \ref{abscoeff}. 
The satellite sequences in these spectra are similar to those found in
the case of the Cs principal series lines \cite{Heinke1983}. 
As seen from Fig. \ref{abscoeff}, the shapes of
higher 5S--nP lines become simpler. They are characterized by single sharp
blue satellites and weak red shoulders. Because of their concise line shapes
they seem to be appropriate candidates for the comparison of experiment and
theory. Thus, our former calculations were extended from the high Rb Rydberg
states around n $\sim$ 30 \cite{Greene2000,Granger2001,Hamilton2002} to the
regime of lower-lying levels, from 9 P to 12 P.

Previous studies of ultra-long-range Rb dimers have suggested that such
extrema arise from two sources. First, the potential of the perturber
interacting with any Rydberg state, even hydrogenic, gives rise to
oscillatory features that reflect the underlying nodal structure of the
Rydberg electron wavefunction. Second, for a nonhydrogenic molecule, the
breaking of the degeneracy of a single n-manifold due to the large quantum
defects of the lower-$l$ states places their energies at points intermediate
to the energies of the high-$l$ degenerate manifolds. Since the molecular
Rydberg state includes some components of these states, the adiabatic
potential can undergo avoided crossings with each of the unperturbed atomic
potentials.

The short-ranged interaction of the Rydberg electron with the ground state perturber atom
approaches a polarization potential far from the
atom, and its center lies far from the ionic core where the Coulomb potential
varies slowly with the radial electron-ion coordinate $r$. This
allows the electron-perturber interaction to be approximated as the scattering of a
quasi-free electron of fixed local momentum from an atom. Far from
the ion, a Rydberg electron has little kinetic energy. Importantly, the low
energy scattering of a $p$-wave electron from a neutral Rb atom displays a
sharp triplet shape resonance.\cite{Hamilton2002,Chibisov2002} 
It is this resonance that spawns avoided
crossings and other extrema near the atomic nP absorption 
lines, and whose electronic structure resembles a butterfly when plotted 
as in Fig.3 of \cite{Hamilton2002}.  Note that the previously predicted 
\cite{Greene2000,Granger2001} 
trilobite states apparently have too 
little atomic P-character to be visible in line-broadening experiments 
that excite directly out of the alkali atom ground states.

To perform a full \textit{ab initio} calculation on the Rydberg electron
states of such a molecule would be exceedingly difficult using existing
quantum chemistry programs, due to the large number of coupled Rydberg
states that contribute to the perturbed wavefunction and the large electron
distances that are relevant. Through most of its range, however, the Rydberg
electron is subject only to the Coulomb potential of the ionic core,
modified by $l$-dependent quantum defects that cause the electron de Broglie
waves to be phase shifted.\cite{SeatonReview} Only in the region of the
perturber does the potential become more complicated. This has given
reasonably accurate results using a zero-range approximation to the
potential, as was generalized to higher-$l$ scattering from the Fermi
pseudopotential by Omont \cite{Omont1977}. This method, however, suffers
from instabilities related to divergent behavior close to a resonance
energy, and thus becomes inaccurate in the regions of the potential giving
rise to avoided crossings.

An alternative method that takes similar advantage of the simplicity of the
potential away from the perturber utilizes the Coulomb Green's function,
which gives an analytical solution to the wavefunction via the Kirchhoff
integral method or similar techniques \cite{Khuskivadze2002,Hamilton2002}.
First the formal solution is written in terms of the Coulomb Green's
function $G_{c}^{mod}(\vec{r},\vec{r}\,^{\prime })$, with appropriate
modifying terms added to correct for the atomic Rydberg state quantum
defects \cite{Khuskivadze2002}.
%
%
Application of Green's theorem converts the resulting expression into a surface
integral over a sphere enclosing the perturber, which implies that we can 
interpret the ground state
perturber as simply modifying the boundary conditions of the 
solution [for details see Eq.14 of \cite{Chibisov2002}, or \cite{HamiltonThesis2002}].
%
%
The matching surface must be sufficiently large that the
scattering from the perturber can be described at that radius entirely in terms
of the asymptotic phase shifts. 
Next the solution is expanded in partial waves at the boundary of the
surface of integration using the spherical Bessel functions,
\begin{equation}
\Psi (\vec{r})=\sum_{lm}A_{lm}\left[ j_{l}(kx)\cos \delta _{l}(\varepsilon
)-\eta _{l}(kx)\sin \delta _{l}(\varepsilon )\right] Y_{lm}(\theta _{x},\phi
_{x}).  \label{eq:InnerSolution}
\end{equation}
\noindent The effect of the perturber interaction is all
contained in the phase shifts $\delta _{l}^{(k)}$. Equating \ref
{eq:InnerSolution} with the Kirchoff integral
and projecting onto the
component spherical harmonics gives a linear system of equations for the
unknown coefficients $A_{lm}$. The bound state energies are
associated with zeroes of a determinantal equation. 

\begin{figure}[h]
\begin{center}
\includegraphics[width=3.7in]{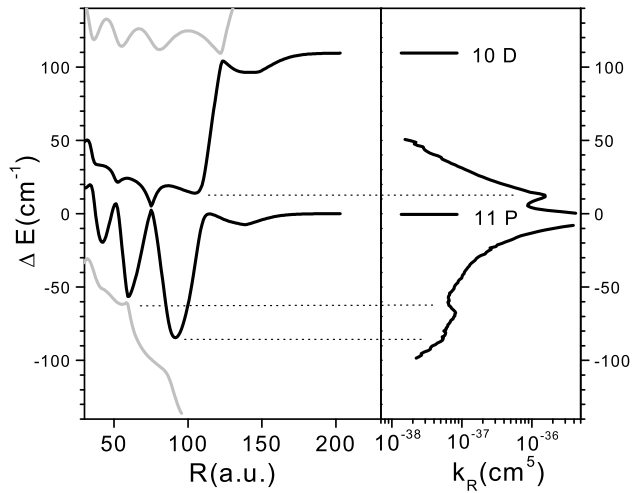}
\end{center}
\caption{$^3\Sigma$ Born-Oppenheimer difference potential curves in the vicinity of the
Rb 11 P level (left) compared with the measured reduced absorption
coefficient of the Rb 5S--11P line (right). The upper and lower gray
potential curves are connected with the 12 S and with
the hydrogenic $n, l$ states having $l\ge 3$,
respectively. The strong blue satellite corresponds to the minimum of the
potential curve that correlates adiabatically with the 10 D level, the 
minimum originating from an avoided crossing between the steeply rising 
``butterfly'' potential curve \cite{Hamilton2002} and the 11 P curve.  
The weak red-shifted shoulders 
correspond to minima in the potential curve that dissociates adiabatically
to the 11 P asymptote, but whose electronic character near the minima is also 
predominantly ``butterfly-like''.  
}
\label{diffpot}
\end{figure}

A detailed view of one absorption line is presented 
in Fig \ref{diffpot}, where the
calculated difference potential curves around the Rb 11P level are compared
with the measured shape of the 5S--11P line. In the case of a single potential
curve $V(R)$ the quasistatic formula for the reduced absorption coefficient
is $
k_{R}=\frac{4\pi ^{2}e^{2}h}{mc}f(R)R^{2}/ \left| \mathrm{d}U(R)/%
\mathrm{d}R\right| .
$
At internuclear distances of interest the ground
state potential is predominantly given by the van der Waals 
interaction\cite{Ban2004}.  According to data taken from
\cite{Klausen2001}, the corresponding $C_6$ coefficient
is 4660 a.u., which, e.g., yields an interaction energy 
$\Delta E_0 \approx -0.065$ cm$^{-1}$ at $R=50$ a.u.
Therefore, the difference potential corresponds in 
practice to the excited state potential alone here.
To calculate the theoretical line shape of $k_R$, the 
$R$-dependent oscillator strengths $f(R)$ are needed,
but they have not yet been calculated in this study.
Nevertheless, the correspondence between the extrema
of the calculated potential curves and the satellite
positions suggest that theory and experiment generally
agree. 
This agreement for a range of $n$%
-values seems unlikely to be fortuitous. Hence we regard this
as a confirmation of the existence of oscillatory long-range potential
curves in these Rydberg-ground state molecule systems.

\begin{figure}[h]
\begin{center}
\includegraphics[width=3.7in]{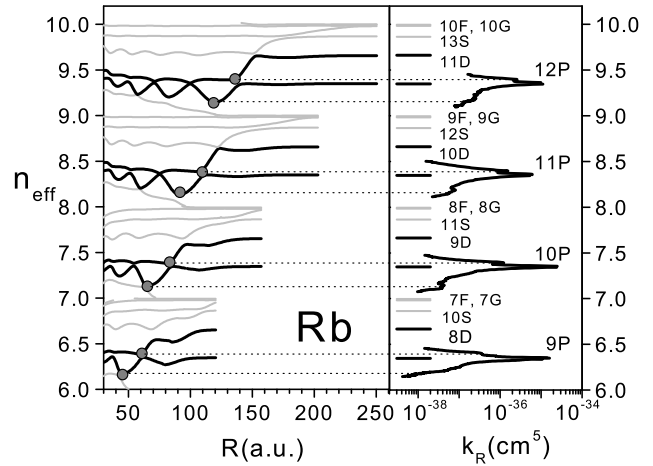}
\end{center}
\caption{Calculated $^3\Sigma$ Rb$^*$-Rb Born-Oppenheimer 
excited state potential curves (left) and
the relevant experimental line shapes (right) of the Rb 5S--nP (n=9, 10, 11,
12) transitions. In contrast to Fig.\ref{diffpot}, the energies are represented 
in terms of the effective principal quantum numbers 
$n_{\textrm{eff}}$. The positions of the
potential curves minima which produce the line satellites are labeled by
gray circles.  
}
\label{effQnum}
\end{figure}

The general agreement between calculated potential curves and measured
absorption profiles is illustrated in 
(Fig.\ref{effQnum}.) Here 
the energy scale is converted to an effective quantum number scale
using $E_n + \Delta E - E_0 = E_i - R_{\textrm Rb}/{n_{\textrm{eff}}}^2,$
where $E_i$ and $R_{\textrm Rb}$ are the ionization limit and the 
Rydberg constant for Rb.  On this scale, 
atomic and molecular levels with successive
principal quantum numbers $n$ are approximately equidistant, which
provides insight into the theoretical and experimental regularities. 
The $R$-values of the deepest potential
minima expressed in Bohr radii scale roughly with $1.5{n_{\textrm{eff}}}^{2}$,
as is expected for Rydberg state length scales. The positions of the line
shoulders on the red line wings and the corresponding potential minima for
the sequence are nearly constant ($\Delta {n_{\textrm{eff}}}=0.25$). This
means that an extrapolation of the experimental red satellites to the
5S--30P line would reveal a potential depth of the Rb*(32P)--Rb(5S)
molecular state comparable to what was 
predicted ($\approx$30 GHz) in \cite{Hamilton2002}.

Preliminary absorption measurements of the self-broadening of the 
K principal series lines revealed very similar satellite structures 
as were found in Rb, although the strengths of the features were 
somewhat weaker.  Thus the Cs$^{\ast }$--Cs case, which displays 
the richest structure of multiple experimental satellites 
\cite{Heinke1983}, appears to be the best example to study in 
order to refine the theory.  However, it is necessary to include 
the spin-orbit coupling, much stronger than in Rb, 
because it has a nontrivial effect on the fine-structure 
intensity ratios, especially in the higher principal series 
lines (see, e.g. \cite{Norcross1973}).

In summary, we have presented experimental evidence for the phenomenon of
oscillatory long-range potential curves, having the qualitative behavior
that was predicted in Refs.\cite{Greene2000,Hamilton2002,Khuskivadze2002}. \
Moreover, these unusual long-range Rydberg molecules provide an explanation
for the origin of certain features that have been observed in
line-broadening experiments for decades, but not successfully interpreted
until now. \ To make the correspondence between theory and experiment
clearer, it will be desirable to improve the Fermi-style zero-range
description of the electron-perturber interaction potential. \ Because lower
principal quantum numbers $n$ are considered in the present experiment and
theoretical analysis, higher electron-perturber collision energies occur
than were important for the $n\sim 30$ \ states studied in Ref.\cite
{Greene2000}. \ This increased collision energy makes the zero-range theory
less applicable for high precision quantitative purposes, but it
nevertheless seems sufficient for the key points of the present
interpretation. \ In the future it will be desirable to calculate the dipole
matrix elements and to thereby hopefully achieve an understanding of the
experimental spectra, at an improved level. 
Also interesting are K$^*$-K, which
looks quite similar to Rb$^*$-Rb, and Cs$^*$-Cs, which shows many more
oscillations and will stringently test theory because spin-orbit
interactions are crucial. \ Experimentally, it is desirable to 
study higher principal quantum numbers, preferably
in a multi-step excitation that might permit the ``trilobite'' 
states to become observable.  Another experimental test of the 
ideas in \cite{Hamilton2002} will be to replace the perturbing ground state species
by a molecule with electron scattering shape resonances, e.g., SF$_{6}$
or CO$_{2}$. 

This work was supported in part by the National Science Foundation and 
the Deutsche Forschungsgemeinschaft. \ We
thank I. I. Fabrikant for helpful correspondence, 
and A. Gallagher for helpful discussions.


\end{document}